\documentclass{article}
\usepackage{amsfonts}
\usepackage{amssymb}
\usepackage{amsmath}
\usepackage{amsthm}
\usepackage{color}
\usepackage[utf8]{inputenc}
\usepackage{comment}

\begin{document}

\title{\bf Cosmological perturbations in modified teleparallel gravity models}
\author{Alexey Golovnev${}^{1}$,  Tomi Koivisto${}^{2,3}$\\
{\small ${}^{1}${\it Faculty of Physics, St. Petersburg State University,}}\\ 
{\small\it Ulyanovskaya ul., d. 1, Saint Petersburg 198504, Russia,}\\
{\small agolovnev@yandex.ru}\\
{\small ${}^{2}${\it Nordita, KTH Royal Institute of Technology and Stockholm University,}}\\
{\small\it Roslagstullsbacken 23, SE-10691 Stockholm, Sweden,}\\
{\small ${}^{3}$ {\it National Institute of Chemical Physics and Biophysics, R\"avala pst. 10, 10143 Tallinn, Estonia,}} \\
{\small tomik@astro.uio.no }
}
\date{}
\hspace{5.2in} \mbox{NORDITA 2018-073} 
{\let\newpage\relax\maketitle}

\begin{abstract}

Cosmological perturbations are considered in $f(T)$ and in scalar-torsion $f(\varphi)T$ teleparallel models of gravity. Full sets of linear perturbation equations are accurately derived and analysed at the relevant limits. Interesting features of generalisations to other teleparallel models, spatially curved backgrounds, and rotated tetrads are pointed out.

\end{abstract}

\section{Introduction}

Teleparallel gravity in its initial formulation is an equivalent way to describe general relativity (GR) in terms of torsion instead of curvature \cite{Aldrovandi:2013wha}. A. Einstein's attempts to incorporate electromagnetism in the reformulation, in spaces characterised by parallelism at distance and originally envisaged by E. Cartan, were unsuccessfull \cite{Goenner:2004se}, but concrete advantages of the reformulation were later clarified by C. M{\o}ller who found a covariant gravitational energy-momentum complex \cite{MOLLER1961118}
and discussed the possible resolution of singularities in the tetrad framework\footnote{One may remind that rather than something ''extra'' or ''alternative'', the tetrad is {\it necessary} to couple matter to gravitation. Only currently, a new preprint claims to consistently supersede the tetrad by a more minimal structure, which may shed new light on the singularity resolution and, surprisingly, the nature of dark matter \cite{Zlosnik:2018qvg}.}
 \cite{moller1978crisis}. In the last decade we have a great resurgence of interest towards such alternative formulations. 
 The reasons range from the need to revisit the foundations of GR \cite{Koivisto:2018aip} to
the phenomenological interests in new approaches to modify gravity \cite{Cai:2015emx}. 
 
In this paper we take the latter viewpoint. In modern cosmology it has become commonly accepted that the existence of the three well-known unknowns, the agents that are supposed to cause inflation, dark matter and dark energy, could serve as a good motivation for modifying the gravitational interaction. In particular, the problems of the early Universe \cite{Ferraro:2006jd,Ferraro:2008ey} and the search for a self-acceleration mechanism in the present day Universe \cite{Bengochea:2008gz,Linder:2010py} have led to one of the simplest modified teleparallel models, the $f(T)$. Since then, many variations of this model have been proposed to the same aims, with (e.g.
\cite{Geng:2011aj,Jarv:2015odu,Hohmann:2018rwf}) and without (e.g. \cite{Kofinas:2014owa,Bahamonde:2015zma,Bahamonde:2017wwk}) additional scalar fields in the action (see \cite{Cai:2015emx} and Section \ref{generalisations}). 
Obviously, a model which aims at solving any of the mentioned cosmological puzzles must be tested against all available sets of cosmological data \cite{Cardone:2012xq}. One of the first steps to be taken is the theory of linear cosmological perturbations.

We will take the classical route of working with zero spin connection in $f(T)$ (pure tetrad formulation) because the covariantisation \cite{Krssak:2015oua,Golovnev:2017dox}, though important at the foundational level, introduces new variables without changing physical predictions which is impractical for our current purposes. The cosmological perturbation equations for $f(T)$ were given first in the Ref. \cite{Li:2011wu}. It was done in the covariant 1+3 language which might be unfamiliar for many workers in the field, and only in the Appendix were they specified to the Newtonian gauge\footnote{Unfortunately, with a typo (missing prime for the extra perturbation in the space-space field equation) inherited in \cite{Nunes:2018xbm}.}. Recently these equations were used \cite{Nunes:2018xbm} in the first actual comparison with the data.
Note that, working with pure tetrad formalism, it is very important to consider the most general perturbation of the tetrad and to take care of all equations of motion including the antisymmetric part self-consistently. Many works\footnote{Both before and after the correct derivations of the linear field equations \cite{Li:2011wu} and the quadratic action \cite{Izumi:2012qj}.} in the field lack consistency in these and other respects, e.g. \cite{Li:2018ixg}. 
We feel it very timely to give a detailed account of cosmological perturbation analysis, with accurate and consistent derivations directly in the Newtonian gauge, in a formalism that is straightforward to generalise to the many variations of the $f(T)$ model.
 
In the next Section \ref{ft} we briefly review the $f(T)$ model, emphasising the crucial point of the presence of antisymmetric field equations which is generic to teleparallel modified gravity models. 
The following Section \ref{parameterisation} introduces a parameterisation of the tetrad perturbations and their gauge transformations. These considerations are totally independent of the particular
model under study ($f(T)$ or otherwise). We shall then derive the perturbed components of torsion and apply them to check the $f(T)$ model sector by sector: tensors, vectors, pseudoscalars and scalars in Sections \ref{tensor}, \ref{vector}, \ref{pseudoscalar} and \ref{scalar}, respectively. The latter are the most non-trivial and the most interesting ones. We apply these scalar perturbations
in Section \ref{comoving} to study the structure formation in the presence of arbitrary matter sources. Generalised models are then discussed in Section \ref{generalisations}, where a more complete study is presented for the case of the scalar-torsion model. Finally, we briefly comment on the generalisation to curved cosmology in Section \ref{nonflat} and point out the issue of inequivalent  choices of ''good'' tetrads. In the conclusions of Section \ref{conclusions}, we list the new results obtained along the way.
 
\section{Brief introduction to $f(T)$}
\label{ft}

The action of $f(T)$ gravity is given in terms of a tetrad field\footnote{The Palatini formulation of teleparallel gravity theories, i.e. the formulation in terms of the metric and an independent affine connection, is also possible and was presented in \cite{BeltranJimenez:2017tkd}.} 
$e^A_{\mu}$, such that $g_{\mu\nu}=\eta_{AB}e^A_{\mu}e^B_{\nu}$, and a non-linear function of the torsion scalar which enters the action of the teleparallel equivalent of GR \cite{Aldrovandi:2013wha}
\begin{equation}
\label{action}
S=-\int d^4 x \| e\|\cdot f\left({\mathbb T}\right)
\end{equation}
where the torsion scalar can be written in several equivalent ways
\begin{eqnarray*}
{\mathbb T} & = &  \frac12 K_{\alpha\beta\mu}T^{\beta\alpha\mu}-T_{\mu}T^{\mu}\\
{} & =  & \frac12 T_{\alpha\beta\mu}S^{\alpha\beta\mu}\\
{} & = &  \frac14 T_{\alpha\beta\mu}T^{\alpha\beta\mu}+\frac12 T_{\alpha\beta\mu}T^{\beta\alpha\mu}-T_{\mu}T^{\mu}
\end{eqnarray*}
with the contortion tensor $K_{\alpha\mu\beta}=-K_{\beta\mu\alpha}$ defined as a difference between the Weitzenb{\"o}ck connection $\Gamma^{\alpha}_{\mu\nu}= e_A^{\alpha}\partial_{\mu}e^A_{\nu}$ and the Levi-Civita connection $\mathop\Gamma\limits^{(0)}{\vphantom{\Gamma}}^{\alpha}_{\mu\nu}$,
\begin{equation*}
\Gamma^{\alpha}_{\mu\nu}=\mathop\Gamma\limits^{(0)}{\vphantom{\Gamma}}^{\alpha}_{\mu\nu}(g)+K^{\alpha}_{\hphantom{\alpha}\mu\nu}
\end{equation*}
which gives
\begin{equation*}
K_{\alpha\mu\nu}=\frac12\left(T_{\alpha\mu\nu}+T_{\nu\alpha\mu}+T_{\mu\alpha\nu}\right)=\frac12\left(T_{\mu\alpha\nu}+T_{\nu\alpha\mu}-T_{\alpha\nu\mu}\right),
\end{equation*}
and with the superpotential
\begin{equation*}
\label{superpot}
S^{\alpha\mu\nu}\equiv K^{\mu\alpha\nu}+g^{\alpha\mu}T^{\nu}-g^{\alpha\nu}T^{\mu}
\end{equation*}
which satisfies the antisymmetry condition $S^{\alpha\mu\nu}=-S^{\alpha\nu\mu}.$ 

One can derive equations of motion by using the following trick. We have the basic relation
\begin{equation*}
\mathop{R}\limits^{(0)}=-{\mathbb T}-2\mathop{\bigtriangledown_{\mu}}\limits^{(0)}T^{\mu}
\end{equation*}
which gives
$$\delta S=-\int\left(f\delta \| e\|+\| e\| f_{T}\delta{\mathbb T}\right)=-\int\left(f\delta \| e\|-\| e\|f_{T}\left(\delta\mathop{R}\limits^{(0)}+2\mathop{\bigtriangledown_{\mu}}\limits^{(0)}\delta T^{\mu}+2T^{\mu}\cdot\delta\left(\frac{1}{\| e\|}\partial_{\mu}\| e\|\right)\right) \right)$$
where the last term comes from varying the notion of the Levi-Civita-covariant derivative (in our paper \cite{Golovnev:2017dox} this term has been missed leading to incorrect symmetric part of equations).

The resulting equation of motion for $f(T)$ gravity in vacuum is
\begin{equation*}
f_{T}\mathop{G_{\mu\nu}}\limits^{(0)}+f_{TT}S_{\mu\nu\alpha}\partial^{\alpha}{\mathbb T}+\frac12 \left(f-f_{T}{\mathbb T}\right)g_{\mu\nu}=0.
\end{equation*}
Interestingly, this nice form of equation can be found already in Ref. \cite{Li:2011wu}. This is our starting point.
Now we add some matter content in terms of the energy-momentun tensor $\Theta_{\mu\nu}$ and use the equation
\begin{equation}
\label{eom}
f_{T}\mathop{G_{\mu\nu}}\limits^{(0)}+f_{TT}S_{\mu\nu\alpha}\partial^{\alpha}{\mathbb T}+\frac12 \left(f-f_{T}{\mathbb T}\right)g_{\mu\nu}=8\pi G\cdot\Theta_{\mu\nu}
\end{equation}
where the variations of the Einstein tensor are well-known from the standard cosmological perturbation theory.

\subsection{Symmetric and antisymmetric parts\\
of equations of motion}

In modified teleparallel gravity models, it is very useful to separate symmetric and antisymmetric parts of equations \cite{Hayashi:1979qx}. The latter are responsible for preferred frame effects, and do coincide with those coming from variation with respect to the flat spin connection \cite{Golovnev:2017dox} in covariantised versions.

Let us first look at the antisymmetric part of equations (\ref{eom}). Suppose that $f_{TT}\neq 0$ and that the energy-momentum tensor of matter is symmetric. In this case we have
\begin{equation*}
\left(S_{\mu\nu\alpha}-S_{\nu\mu\alpha}\right)\partial^{\alpha}{\mathbb T}=0.
\end{equation*}
One can easily see that it boils down to 
\begin{equation}
\label{aseom}
\left(T_{\alpha\mu\nu}+g_{\alpha\mu}T_{\nu}-g_{\alpha\nu}T_{\mu}\right)\partial^{\alpha}{\mathbb T}=0.
\end{equation}

We note in passing that, interestingly enough, while perturbation  with respect to an arbitrary spin connection makes the expression in brackets vanish (and therefore zero torsion), see Ref. \cite{Golovnev:2017dox}, variation in inertial class demands it be orthogonal to $\partial_{\alpha}{\mathbb T}$. 

For symmetric part of equations, let us denote
\begin{equation}
\label{Q}
Q_{\mu\nu}\equiv\frac12 \left(S_{\mu\nu\alpha}+S_{\nu\mu\alpha}\right)\partial^{\alpha}\mathbb T,
\end{equation}
where we easily see that
\begin{equation*}
S_{\mu\nu\alpha}+S_{\nu\mu\alpha}=T_{\mu\nu\alpha}+T_{\nu\mu\alpha}+2g_{\mu\nu}T_{\alpha}-\left(g_{\alpha\mu}T_{\nu}+g_{\alpha\nu}T_{\mu}\right),
\end{equation*}
and write the symmetric part of equation (\ref{eom}) as
\begin{equation}
\label{seom}
f_{T}\mathop{G^{\mu}_{\nu}}\limits^{(0)}+f_{TT}Q^{\mu}_{\nu}+\frac12 \left(f-f_{T}{\mathbb T}\right)\delta^{\mu}_{\nu}=8\pi G \Theta^{\mu}_{\nu}.
\end{equation}

Let us also make a general comment on the structure of equations. The broken local Lorentz invariance implies that the Bianchi identities do not hold automatically. Indeed, if we define ${\mathfrak T}^{\mu\nu}$ via
$$\frac{\delta S}{\delta e^A_{\mu}}\equiv \| e\| {\mathfrak T}^{\mu\nu} e^B_{\nu}\eta_{AB},$$
then invariance of the action under diffeomorphisms $e^A_{\mu}\longrightarrow e^A_{\mu}-e^A_{\nu}\partial_{\mu}\zeta^{\nu}-\zeta^{\nu}\partial_{\nu}e^A_{\mu}$ leads to
$$\frac{1}{\| e\|}\partial_{\mu}\left(\| e\|{\mathfrak T}^{\mu}_{\nu}\right)-{\mathfrak T}^{\beta}_{\alpha}e^{\alpha}_{A}\partial_{\nu}e^A_{\beta}=0$$
which can easily be transformed (using $K_{\alpha\mu\beta}-T_{\alpha\mu\beta}=-K_{\mu\beta\alpha}$) into
$$\mathop{\bigtriangledown_{\mu}}\limits^{(0)}{\mathfrak T}^{\mu\nu}+K^{\alpha\nu\beta}{\mathfrak T}_{\alpha\beta}=0.$$

When the local Lorentz invariance is satisfied, invariance under $e^a_{\mu}\longrightarrow \Lambda^a_b e^b_{\mu}$ implies that ${\mathfrak T}^{\mu\nu}$ is symmetric, and by virtue of antisymmetry of contortion tensor, the usual Bianchi identities are restored. In $f(T)$ this is not the case. However, the antisymmetric part of equations requires that the antisymmetric part of ${\mathfrak T}^{\mu\nu}$ vanishes, and after that the Bianchi identities are in operation again. Apparently, cases with fermions and/or spin-density should be considered with care\footnote{It has been recently clarified that, in contrast to the present framework, the natural coupling prescription is consistent and viable in symmetric teleparallelism \cite{BeltranJimenez:2017tkd,Koivisto:2018aip}. This can be considered as a criterion that determines the real ''physical'' geometry \cite{So:2006pm} in favour of the symmetric teleparallel theory; for further arguments see \cite{BeltranJimenez:2017tkd,Koivisto:2018aip}.}.

\subsection{Spatially flat FRW cosmology}

We consider the FRW (Friedmann-Robertson-Walker) cosmology
$$ds^2=a^2(\tau)\left(-d\tau^2+dx^i dx^i\right)$$
in terms of the following tetrad ansatz:
$$e^A_{\mu}=a(\tau)\cdot\delta^A_{\mu}.$$
We note that a tetrad choice for the given metric has been done which is not innocuous in the context of $f(T)$, but in this case seems reasonable.

Moreover, this choice corresponds to what is known as a "good" tetrad \cite{Tamanini:2012hg} (we shall return to this important and subtle issue in Section \ref{nonflat}). Indeed, in general the equations are symmetric either if in teleparallel equivalent of GR ($f_{TT}= 0$) or for constant torsion scalar solutions ($\partial_{\alpha}{\mathbb T}=0$). However, the FRW ansatz goes through eq. (\ref{aseom}) because the background tensor $S_{\mu\nu\alpha}-S_{\nu\mu\alpha}$ has non-zero components only for spatial values of index $\alpha$,
\begin{equation*}
S_{i0j}=-S_{ij0}=-2a^2 H\delta_{ij}\,,
\end{equation*}
 while the torsion scalar $\mathbb T$ depends only on time.

Symmetric part of equations is quite standard with a new piece given by the tensor $Q$ above (\ref{Q}). At the background level the only non-trivial components of $Q$ are spatial
$$Q^i_j=-\frac{24H^2}{a^4}(H^{\prime} - H^2)\delta^i_j\,.$$

For the torsion vector and torsion scalar we have
$$T_{\mu}=3H\delta^0_{\mu}\,,$$
$${\mathbb T}\equiv\frac12 S^{\alpha\mu\nu}T_{\alpha\mu\nu}=\frac{6}{a^2}H^2\,.$$
For a systematic and exhaustive analysis of the background dynamics, we refer to \cite{Hohmann:2017jao}.

\section{Tetrad perturbation and gauge freedom}
\label{parameterisation}

One can parametrise linear perturbations by
\begin{eqnarray*}
e^{\emptyset}_0 & = & a(\tau)\cdot\left(1+\phi\right)\\
e^{\emptyset}_i & = & a(\tau)\cdot\left( \partial_i \beta+u_i\right)\\
e^a_0 & = & a(\tau)\cdot\left( \partial_a \zeta+v_a\right)\\
e^a_j & = & a(\tau)\cdot \left((1-\psi)\delta^a_j+\partial^2_{aj}\sigma+\epsilon_{ajk}\partial_k s+\partial_j c_a+\epsilon_{ajk}w_k+\frac12 h_{aj}\right).
\end{eqnarray*}
where as usual all vectors are assumed to be divergenceless, and the tensorial part is also traceless. Note that we do not symmetrise in  the $\partial c$-part of the tetrad since this choice allows for a simple description of infinitesimal diffeomorphisms, see below.

The corresponding metric components are given by
\begin{eqnarray*}
g_{00} & = & -a^2(\tau)\cdot\left(1+2\phi\right)\\
g_{0i} & = & a^2(\tau)\cdot\left( \partial_i \left(\zeta-\beta\right)+v_i-u_i\right)\\
g_{ij} & = & a^2(\tau)\cdot \left((1-2\psi)\delta_{ij}+2\partial^2_{ij}\sigma++\partial_i c_j + \partial_j c_i +h_{ij}\right).
\end{eqnarray*}
Compared to GR, we have new components given by scalar $\beta+\zeta$, pseudoscalar $s$, vector $u_i+v_i$, and pseudovector $w_j$. Those 6 variables correspond to local Lorentz rotations of the tetrad: 3 boosts in direction of $u+v+\partial ({\beta}+\zeta)$ and 3 rotations around $w+\partial s$.

Under infinitesimal diffeomorphisms
$x^{\mu}\to x^{\mu}+\xi^{\mu}(x)$
with $\xi^0$ the time-component and $\xi^i\equiv\partial_i\xi+{\tilde\xi}_i$ the spatial components, where ${\tilde\xi}_i$ is transverse, one can simply derive the following transformation laws:
\begin{eqnarray*}
\phi & \longrightarrow & \phi-{\xi^0}^{\prime}-H\xi^0\\
\psi & \longrightarrow & \psi+H\xi^0\\
\sigma & \longrightarrow & \sigma-\xi\\
\beta & \longrightarrow & \beta-\xi^0\\
\zeta & \longrightarrow & \zeta-\xi^{\prime}\\
c_i & \longrightarrow & c_i -{\tilde\xi}_i\\
v_i & \longrightarrow & v_i -{{\tilde\xi}_i}^{\prime}.
\end{eqnarray*}
Gauge invariant combinations are obvious. For the subsequent calculations, our choice would be  $\sigma=0$ and $\beta=\zeta$ (conformal Newtonian gauge), and $c_i=0$.

\section{Tensor perturbations}
\label{tensor}

Let us start from the simplest part of transverse traceless perturbations.
It is easy to see that, in this sector, the only non-zero torsion components are
$$T_{ijk}=\frac{a^2}{2}\left(\partial_j h_{ik}-\partial_k h_{ij}\right),$$
$$T_{i0j}=a^2\left(H\delta_{ij}+\frac12\left(h_{ij}^{\prime}+2Hh_{ij}\right)\right),$$
and both $\delta\mathbb T=0$ and $\delta T_{\mu}=0$.

The antisymmetric part of equation (\ref{eom}) is satisfied identically. In the symmetric part (\ref{seom}) we have
$$Q^{i}_{j}=\left( T^i_{\hphantom{i}j0} + \delta^i_j T_0\right) \partial^0{\mathbb T}=\frac{12H(H^{\prime}-H^2)}{a^4}\left(-2H\delta_{ij}+\frac12 h_{ij}^{\prime}\right)$$
which via the standard relation
$$a^2\delta G^i_j=\frac12\left(h_{ij}^{\prime\prime}+2Hh_{ij}^{\prime}-\bigtriangleup h_{ij}\right)$$ leads to
$$f_Th_{ij}^{\prime\prime}+2H\left(f_T + \frac{6f_{TT}(H^{\prime}-H^2)}{a^2}\right)h_{ij}^{\prime}-f_T \bigtriangleup h_{ij}=0$$
for an ideal fluid. This is very similar to the usual propagation of the gravitational waves. 
Since merely the unobservable effect of the Hubble friction is now slightly modified, 
we immediately see that the models are not constrained \cite{Cai:2018rzd} by the current gravitational wave data. 

\section{Vector (and pseudovector) perturbations}
\label{vector}

For the vector and pseudovector perturbations we use the gauge freedom to set $c=0$ and easily compute the following torsion components:
\begin{eqnarray*}
T_{0ij} & = & a^2\left(\partial_j u_i-\partial_i u_j\right)\\
T_{00i} & = & a^2\partial_i\left(-u_i^{\prime}+H(v_i-u_i)\right)\\
T_{ijk} & = & a^2\cdot \left(\epsilon_{ikl}\partial_j w_l-\epsilon_{ijl}\partial_k w_l\right)\\
T_{i0j} & = & a^2\left(H\delta_{ij}+\epsilon_{ijk}w_k^{\prime}-\partial_j v_i\right)
\end{eqnarray*}
and the first order variation in the torsion vector given by
$$T_i=\epsilon_{ijk}\partial_j w_k$$
while $T_0$ and $\mathbb T$ receive no linear variation.

The antisymmetric part of equations $\left(T_{\alpha\mu\nu}+g_{\alpha\mu}T_{\nu}-g_{\alpha\nu}T_{\mu}\right)\partial^{\alpha}{\mathbb T}=0$ then boils down to
$$T_{0\mu\nu}+g_{0\mu}T_{\nu}-g_{0\nu}T_{\mu}=0.$$
With spatial indices we get
$$\partial_j u_i-\partial_i u_j=0$$
which, after taking divergence, implies $\bigtriangleup u_i=0$ and, in perturbation theory, should be solved as $$u_i=0.$$ 

The mixed indices case gives
$$u_i^{\prime}+2H(v_i-u_i)+\epsilon_{ijk}\partial_j w_k=0$$
which, given that $u=0$, constrains $w$ as $\epsilon_{ijk}\partial_j w_k=-2Hv_i$. Here we have two independent equations for two independent components of $w$. 

Now let us look at the symmetric part (\ref{seom}).
For mixed indices we easily find that
$$Q_{0i}=-\frac{6H(H^{\prime}-H^2)}{a^2}\left(u_i^{\prime}+2H(v_i-u_i)+\epsilon_{ijk}\partial_j w_k\right)$$
which vanishes under the antisymmetric part of equations. Therefore, this part of Einstein equations is not modified:
$$f_{T}\bigtriangleup v_i=16\pi G a^2 (\rho+p){\mathfrak u}_i$$
where ${\mathfrak u}$ is the vortical part of ideal fluid velocity, and we have used $u=0$ to write simply $v$ instead of the metric perturbation $v-u$.

Analogously we find
$$\delta Q_{ij}=-\frac{6H(H^{\prime}-H^2)}{a^2}\left(\partial_i v_j + \partial_j v_i\right)$$
which with 
$$a^2\delta G^i_j=-\frac12\left(\partial_i v_j + \partial_j v_i\right)^{\prime}-H\left(\partial_i v_j + \partial_j v_i\right)$$ gives
$$f_{T}\cdot v^{\prime}+2\left(f_T H +\frac{6f_{TT}H(H^{\prime}-H^2)}{a^2}\right)v=0$$
in case of perfect fluid matter.

\section{Pseudoscalar perturbation}
\label{pseudoscalar}

Let us consider the pseudoscalar perturbation
$$e^a_i=a(\tau)\left(\delta^a_i+\epsilon_{aij}\partial_j s\right)$$
with
 the only non-zero components of the torsion tensor given by
$$T_{i0j}=-T_{ij0}=a^2\left(H\delta_{ij}+\epsilon_{ijk}\partial_k s^{\prime}\right).$$
It is easy to see that this perturbation does not contribute to the linear perturbation equations at all. One can think of it as a remnant symmetry.

\section{Scalar perturbations (Newtonian gauge)}
\label{scalar}

For the scalars, let us choose the Newtonian gauge as $\sigma=0$ and $\beta=\zeta$.
We get the following linearised torsion components:
\begin{eqnarray*}
T_{0ij} & = & 0\\
T_{00i} & = & a^2\partial_i\left(\phi-\zeta^{\prime}\right)\\
T_{ijk} & = & a^2\cdot \left(\delta_{ij}\partial_k \psi-\delta_{ik}\partial_j\psi\right)\\
T_{i0j} & = & a^2\left[H\delta_{ij}- \partial^2_{ij}\zeta- \delta_{ij}\left(2H\psi+\psi^{\prime}\right)\right].
\end{eqnarray*}
 Note that $T_{i0j}$ is symmetric under $i\leftrightarrow j$.

 Let us also find the torsion vector
\begin{equation*}
T_i =\partial_i \left(\phi-\zeta^{\prime}-2\psi\right),
\end{equation*}
$$T_0=3H-\bigtriangleup\zeta-3\psi^{\prime},$$
and the variation of the torsion scalar
\begin{equation*}
\delta{\mathbb T}=-\frac{4H}{a^2}\left(\bigtriangleup\zeta+3H\phi+3\psi^{\prime}\right).
\end{equation*}

One can easily check that $\left(S_{ij\alpha}-S_{ji\alpha}\right)\partial^{\alpha}{\mathbb T}$ vanishes identically in the linear order. Then, at this order, the antisymmetric variation comes from mixed indices and gives
\begin{equation*}
-S_{i0j}\partial^j{\delta\mathbb T}+\left(\delta S_{0i0}-\delta S_{i00}\right)\partial^0{\mathbb T}=0.
\end{equation*}
We easily find that $$\delta S_{0i0}-\delta S_{i00}=-2a^2\partial_i\psi,$$ and given our expressions for ${\mathbb T}$ and $\delta{\mathbb T}$, the final form of the antisymmetric part of perturbation equation reads
\begin{equation*}
\partial_i\left(H\bigtriangleup\zeta+3H^2\phi+3H\psi^{\prime}-3H^{\prime}\psi+3H^2\psi\right)=0.
\end{equation*}

We can solve this equation as
\begin{equation}
\label{zeta}
\bigtriangleup\zeta=-3\left(\psi^{\prime}+H\phi-\frac{H^{\prime}-H^2}{H}\psi\right).
\end{equation}
and conclude that the antisymmetric part of perturbations makes perfect sense making the (essentially Lorentz) variable $\zeta$ constrained (compare also to vector sector). Now, we have the usual number of equations for the usual number of variables for the symmetric part. 

Symmetric part is a bit more cumbersome. It is easy to see that at the linear level $\delta Q^0_0=0$ and the equation (\ref{seom})
\begin{equation*}
f_{T}\delta\mathop{G^0_0}\limits^{(0)}+f_{TT}\left(\mathop{G^0_0}\limits^{(0)}-\frac12 {\mathbb T}\right)\delta\mathbb T=-8\pi G \delta\rho
\end{equation*}
yields the result
\begin{equation}
\label{time}
f_{T}\bigtriangleup\psi-3H\left(f_{T}+\frac{12H^2}{a^2}f_{TT}\right)\left(\psi^{\prime}+H\phi\right)-12\frac{f_{TT}H^3}{a^2}\bigtriangleup\zeta=4\pi Ga^2\delta\rho
\end{equation}
where we have used
\begin{equation*}
a^2\mathop{G^0_0}\limits^{(0)}=-3H^2-2\bigtriangleup\psi+6H\left(\psi^{\prime} +H\phi\right)\quad {\mathrm{and}}\quad \delta\Theta^0_0=-\delta\rho.
\end{equation*}

For the mixed components we find
$$\delta Q^0_i=-\frac{4H^2}{a^4}\partial_i\left(\bigtriangleup\zeta+3H\phi+3\psi^{\prime}+3\frac{H^{\prime} - H^2}{H}\psi\right)$$
which, with 
$$a^2\delta\mathop{G^0_i}\limits^{(0)}=-2\partial_i\left(\psi^{\prime} +H\phi\right) \quad {\mathrm{and}}\quad  \delta\Theta^0_i=-(\rho+p)\partial_i \mathfrak u,$$ 
brings the equation (\ref{seom})
\begin{equation*}
f_{T}\delta\mathop{G^0_i}\limits^{(0)}+f_{TT}\delta Q^0_i=8\pi G  \delta\Theta^0_i
\end{equation*}
into the form
$$f_{T}\left(\psi^{\prime}+H\phi\right)+\frac{2H^2}{a^2}f_{TT}\left(\bigtriangleup\zeta+3H\phi+3\psi^{\prime}+3\frac{H^{\prime} - H^2}{H}\psi\right)=4\pi Ga^2(\rho+p)\delta \mathfrak u$$
where $\mathfrak u$ is the velocity potential. 

Using our solution (\ref{zeta})  for $\zeta$ the latter equation can be brought to a nicer form of
\begin{equation} \label{eq0i}
f_{T}\left(\psi^{\prime}+H\phi\right)+\frac{12H\left(H^{\prime} - H^2\right)f_{TT}}{a^2}\psi=4\pi Ga^2 (\rho+p)\delta\mathfrak u.
\end{equation}
As usual, it constrains the velocity potential. 

For the spatial part, we compute
$$Q^i_j=\left(-T^i_{\hphantom{i}0j}+\delta^i_j T_0\right)g^{00}\partial_0 \mathbb T$$
at linear order which is a product of two factors:
$$\left(-T^i_{\hphantom{i}0j}+\delta^i_j T_0\right)g^{00}=\frac{1}{a^2}\left(-2H\delta^i_j-\partial^2_{ij}\zeta+\delta^i_j\left(\bigtriangleup\zeta+2\psi^{\prime}+4H\phi\right)\right)$$
and
\begin{equation*}
\partial_0 \mathbb T=\frac{12H(H^{\prime}-H^2)}{a^2}-\frac{4(H^{\prime}-2H^2)}{a^2}\left(\bigtriangleup\zeta+3H\phi+3\psi^{\prime}\right)
-\frac{4H}{a^2}\left(\bigtriangleup\zeta^{\prime}+3H^{\prime}\phi+3H\phi^{\prime}+3\psi^{\prime\prime}\right).
\end{equation*}

This is to be substituted into the perturbed equation (\ref{seom})
\begin{equation*}
f_{T}\delta\mathop{G^i_j}\limits^{(0)}+f_{TT}\delta Q^i_j+\left(f_{TTT}Q^i_j+f_{TT}\left(\mathop{G^i_j}\limits^{(0)}-\frac12 {\mathbb T}\delta^i_j\right)\right)\delta\mathbb T= 8\pi G \delta p \delta^i_j
\end{equation*}
using also expression for $\delta\mathbb T$ and $\mathop{G^i_j}\limits^{(0)}-\frac12 {\mathbb T}\delta^i_j=-\frac{1}{a^2}(2H^{\prime}+4H^2)\delta^i_j$ and
\begin{equation*}
a^2\delta\mathop{G^i_j}\limits^{(0)}=\delta^i_j\left(2\psi^{\prime\prime}+H(4\psi+2\phi)^{\prime}+2(H^2+2H^{\prime})\phi+\bigtriangleup (\phi-\psi)\right)-\partial^2_{ij}(\phi-\psi)
\end{equation*}
and $\delta\Theta^i_j=\delta p\delta^i_j$ for perfect fluid matter.

Assuming no anisotropic stress, the $\partial^2_{ij}$ part of equation gives
$$f_{T}(\phi-\psi)+12f_{TT}H(H^{\prime} - H^2)\zeta=0.$$
It is interesting to note that we have gravitational slip
\begin{equation}
\label{slip}
\phi-\psi=-\frac{12f_{TT}H(H^{\prime} - H^2)}{f_{T}}\zeta
\end{equation}
even without anisotropic stress. Moreover, given our solution (\ref{zeta}) for $\zeta$,
it is a very big slip for superhorizon modes, unless very close to de Sitter.

For the remaining piece of information, we multiply the spatial components of equations by $\frac13 \delta^j_i$ (and then further by $\frac 12$) which after some elementary calculations gives
\begin{multline}
\label{space}
f_{T}\left(\psi^{\prime\prime}+H(2\psi+\phi)^{\prime}+(H^2+2H^{\prime})\phi+\frac13 \bigtriangleup (\phi-\psi)\right)\\
+\frac{4f_{TT}}{a^2}\left(H^2\bigtriangleup\zeta^{\prime}+H(3H^{\prime}-H^2)\bigtriangleup\zeta\right)\\
+\frac{12f_{TT}}{a^2}\left(H^2\psi^{\prime\prime}+H(3H^{\prime}-H^2)\psi^{\prime}+H^3\phi^{\prime}+H^2(5H^{\prime}-2H^2)\phi\right)\\
+\frac{48f_{TTT}H^3(H^{\prime}-H^2)}{a^4}\left(\bigtriangleup\zeta+3H\phi+3\psi^{\prime}\right)=4\pi G a^2 \delta p.
\end{multline}

Precisely as we have done above for the mixed components, one can substitute the solution (\ref{zeta}) for $\bigtriangleup\zeta$ from the antisymmetric part into the temporal (\ref{time}) and diagonal spatial (\ref{space}) components (or even into the intermediate steps of derivations for then the way to them will become much shorter) and get
\begin{equation} \label{eq00}
f_{T}\left(\bigtriangleup\psi-3H(\psi^{\prime}+H\phi)\right)-\frac{36f_{TT}H^2(H^{\prime}-H^2)}{a^2}\psi=4\pi Ga^2\delta\rho
\end{equation}
and 
\begin{multline} \label{eqii}
f_{T}\left(\psi^{\prime\prime}+H(2\psi+\phi)^{\prime}+(H^2+2H^{\prime})\phi+\frac13 \bigtriangleup (\phi-\psi)\right)\\
+\frac{12f_{TT}}{a^2}\left(H(H^{\prime}-H^2)\psi^{\prime}+\left(HH^{\prime\prime}+2{H^{\prime}}^2-5H^2 H^{\prime}+H^4\right)\psi+H^2(H^{\prime}-H^2)\phi\right)\\
+\frac{144f_{TTT}H^2(H^{\prime}-H^2)^2}{a^4}\psi=4\pi G a^2 \delta p,
\end{multline}
which, in a way, look very similar to the GR.

If, for a given mode with a wavenumber $k$, we solve for the gravitational slip as, see eqs. (\ref{zeta}) and (\ref{slip}),
\begin{equation*}
-k^2f_{T}\left(\phi-\psi\right)=36f_{TT}(H^{\prime} - H^2)\left(H\psi^{\prime}+H^2\phi-\left(H^{\prime}-H^2\right)\psi\right),
\end{equation*}
substitute $\phi$ as a function of $\psi$ and combine two equations (\ref{eq00}) and (\ref{eqii}) for an adiabatic mode by $$\delta p=c_s^2 \delta\rho,$$ we will get a second order equation for $\psi$, much the same way as in GR, though not that nice.

Note that in the limit of $k\to\infty$ the gravitational slip vanishes, and as seen in the next Section, the system becomes very tractable.

On the other hand, if $k\to 0$ then either the gravitational slip diverges, or we set $\phi\to -\left(1-\frac{H^{\prime}}{H^2}\right)\psi-\frac{\psi^{\prime}}{H}$ which makes $\psi^{\prime\prime}$ drop out from the equation. In the next Section we however verify that this feature is not a signal of a pathology. Anyway, the prediction of large scale gravitational slip can be tested against cosmic microwave background observations \cite{Nunes:2018xbm} and there is interesting potential in the Square Kilometer Array experiment \cite{Carilli:2004nx} to probe the cosmological structure at the very largest scales especially
via intensity mapping. 
Galaxy clustering and cosmic shear measurements as a means to constrain $f(T)$ models were considered in Ref. \cite{Camera:2013bwa}, which reported a pioneering forecast on the constraints to be expected
from the Euclid data \cite{Amendola:2016saw} on teleparallel models of modified gravity.

\section{Evolution of scalar perturbations (comoving gauge)}
\label{comoving}

It is easy to see that in the absence of matter, the only solution to the system of equations (\ref{eq0i},\ref{eq00},\ref{eqii}) is $\phi=\psi=0$. In vacuum there are no scalar perturbations. It is still not guaranteed that the additional degrees of freedom would not begin to propagate in the presence of matter. For one thing, the vacuum solutions of $f(T)$ cosmology are flat or de Sitter (barring the singular case $f(T) \sim \sqrt{T}$), and the modes could be strongly coupled there. Another thing is their possible excitation by the fluctuations of the matter source. However, K. Izumi and Y.-C. Ong \cite{Izumi:2012qj} had found that in the presence of a canonical scalar field, no additional poles appear in the propagator of the scalar perturbations. We will confirm this by deriving the evolution equation for the single scalar perturbation in the slightly more general setting without any assumptions about the matter source(s).

The energy momentum tensor of generic matter can be parameterised in the fluid form as 
\begin{eqnarray}
\Theta^0{}_0 & = & \rho\left( 1+\delta\right)\,, \nonumber \\ 
\Theta^i{}_0  & = &  \left( \rho + p \right) \delta \mathfrak{u}_{,i}\,, \nonumber \\  
\Theta^i{}_j & = & \delta^i_ j\left( p+\delta p\right) + \rho\left( \Pi_{,ij}-\frac{1}{3}\delta^i_j\bigtriangleup\Pi\right)\,, \label{emt}  
\end{eqnarray}
where $\rho$ is the background energy density and $p$ the background pressure, and $\delta\rho=\rho\delta$ and $\delta p$ are their perturbations. If the fluid is adiabatic,
$\delta p = c_s^2 \delta\rho$, where $c_s^2=p'/\rho'$. More generally, it is conventional to parameterise the pressure perturbation in the comoving gauge, where the velocity perturbation vanishes. We will continue to work with the Newtonian gauge metric perturbations, and denote the comoving gauge matter perturbations with a hat. Thus we have 
\begin{eqnarray}
\hat{\delta \mathfrak{u}} &= &0\,, \\
\hat{\delta} & = & \delta + 3H(1+w)\delta \mathfrak{u}\,, \\
\hat{\delta p} & = & \delta p + 3H(\rho+p)c_s^2\delta \mathfrak{u}\,, \\
\hat{\Pi} & = & \Pi\,.
\end{eqnarray}
Now by combining (\ref{eq0i}) and (\ref{eq00}) we can see that the Poisson equation, which is well known is GR, holds also exactly for linear perturbations in $f(T)$ gravity:
\begin{equation} \label{poisson}
\bigtriangleup \psi = 4\pi G a^2\rho\hat{\delta}\,.
\end{equation} 
It has been had noted previously \cite{Wu:2012hs,Nunes:2018xbm} that the equation is valid for Newtonian gauge perturbations at the small scale limit, which can be understood by that $\delta$ and $\hat{\delta}$ become equal when $k \rightarrow \infty$ since there the velocity perturbation vanishes $\delta\mathfrak{u} \rightarrow 0$. 

As a cross-check of our equations, we use the source (\ref{emt}) in the field equations, and use them (including the derivatives of (\ref{eq0i}) and (\ref{eq00})) to obtain
\begin{equation}
\delta' = (1+w)\left(\bigtriangleup \delta \mathfrak{u}+3\psi'\right) + 3H\left(w\delta - \frac{\delta p}{\rho}\right)\,, \nonumber
\end{equation}
and
\begin{equation}
\delta \mathfrak{u}' = -H\left(1-3w\right)\delta \mathfrak{u} - \frac{w'}{1+w}\delta \mathfrak{u} + \frac{\delta p}{\rho+p} + \phi - \frac{2}{3(1+w)a^2}\Pi \,. \nonumber
\end{equation}
As they should, these correspond precisely to the two non-trivial perturbed components of the conservation equation $\mathop{\nabla}\limits^{(0)}{}_\mu T^\mu{}_\nu=0$. 

As mentioned earlier, we can solve all perturbations in terms of one them, for which we then obtain a closed second order evolution equation. The so called Bardeen equation for $\psi$ is related, via the Poisson equation (\ref{poisson}), to the evolution equation for the comoving density perturbation $\hat{\delta}$, whose spectrum is an important cosmological observable. We report only the result. The equation has the form\footnote{We stress that the equation (\ref{bardeen}) is completely general, and valid even if the matter sector consists of several distinct sources. Of course, one may then have to use additional equations, depending on the specifications of the matter sector, to determine the anisotropic stress $\Pi$ and the isotropic pressure parameter $\hat{c}^2$. Generically for multi-fluid systems $\hat{c}^2_s \neq c_s^2$ even if the individual fluids were adiabatic.}
\begin{equation}
\hat{\delta}'' + A H\hat{\delta}' = \left( B H^2 + \hat{c}_s^2\bigtriangleup\right) \hat{\delta} - \frac{2H}{a^2}\Pi' + \left(C \frac{H^2}{a^2} + \frac{2}{3a^2}\bigtriangleup\right)\Pi\,, \label{bardeen}
\end{equation}
where $\hat{c}_s^2 = \hat{\delta p}/\hat{\delta \rho}$, and the three dimensionless coefficients $A$, $B$, and $C$   depend upon the function $f(T)$. Their general dependence is somewhat messy, and we write these functions down only at the interesting small-scale limit. There
\begin{eqnarray}
A & \rightarrow &  1 + 3c_s^2 - 6w\,, \nonumber \\
B & \rightarrow &  3(5w -3c_s^2 ) - (1+w)\frac{f_T(1-3w)\frac{8\pi G\rho}{H^2} + 12a^{-2} f_{TT}(1+3\hat{c}_s^2)}{2f_T\left( f_T+12f_{TT}H^2\right)}\,, \nonumber \\
C & \rightarrow & 6(w-c_s^2) + \frac{2(1+w)\frac{8\pi G\rho}{H^2}}{f_T + 12a^{-2}f_{TT}H^2}\,. \nonumber
\end{eqnarray}
In the so called quasi-static approximation, we thus obtain for a dust source simply
\begin{equation} \label{delta}
\hat{\delta}'' + H\hat{\delta}' = \frac{4\pi G}{f_T}a^2\rho\hat{\delta}\,.
\end{equation}
In fact, for the comoving overdensity this equation is exact at all scales i.e. without the quasi-static approximation.
Thus the linear structure formation during the matter dominated era is only affected by the modified background evolution through $H$ and the modulation of the effective Newton's constant through $f_{T}^{-1}G$. Note that rather than the Newtonian, the comoving gauge is the best approximation to the ''physical gauge'' wherein we make the measurements of the structure of the Universe around us. 

At large scales, the three functions are a bit more complicated. As an example 
\begin{eqnarray}
A & \rightarrow &  2(1-3w+3c_s^2) \nonumber \\
& + & \frac{8\pi G\rho}{H^2}(1+w)\frac{a^2 f_T^2f_{TT} + 6(f_Tf_{TTT}-2f_{TT}^2)H^2 -72a^{-2}f_{TT}^3H^4}{f_Tf_{TT}H(f_T+12a^{-2} f_{TT} H^2)^2}\,. 
\end{eqnarray}
The two other coefficients $B$ and $C$ are also well-behaved at the homogeneous limit. In particular, they do not contain any divergent terms such as $\sim k^{-2}$. Unlike in GR, the pressure perturbation is relevant even at the homogeneous limit, since now $B$ in general depends also on the $\hat{c}_s^2$ and not only on the $w$ and the ${c}_s^2$ that characterise the background evolution of the fluid. 

\subsection{Bounces}

As another application, let us consider perturbations in bouncing cosmology. Bouncing backgrounds cosmologies have been constructed in teleparallel models 
(see e.g. \cite{Ferraro:2008ey} for an early and \cite{delaCruz-Dombriz:2018nvt} for a recent work), but the problem of perturbations has not been addressed. Non-singular background 
cosmologies often need to resort to matching of the perturbations across the bounce, if not plagued by ghosts or already classical instabilities of linear perturbations \cite{Battefeld:2014uga}.
 
We take the function $f(T)$ to be such that $f_{TT}(0)$ is a finite constant, since it is natural assume analyticity, and this is sufficient for ultraviolet modifications. 
It is also natural to assume that $f_T(0)=1$, so that the modifications are introduced around the standard theory.
At a bounce point of a non-singular cosmology, and at a turnover point of a recollapsing scenario, the evolution equation (\ref{bardeen}) becomes
\begin{equation} \label{bounce}
\hat{\delta}'' =  {4\pi \rho}(1+w)\left[ \left( 1-3w + 2f_{TT}(0)(24\pi G\rho(1+w)\frac{a}{k})^2 \right)\hat{\delta}
+ \frac{4}{a^2}\Pi\right] - k^2\left( \hat{c}_s^2\hat{\delta} - \frac{2}{3a^2}\Pi \right) \,.
\end{equation}
The condition to avoid a divergence of the homogeneous mode of the perturbations is that $f_{TT}(0)=0$, though even then it not entirely 
clear that the homogeneous mode would be well-behaved at a higher order in perturbation theory.
From the (\ref{eom}) it is evident that in a matter bounce, a negative cosmological constant would be needed to cancel the matter energy density at the bounce. 
Since the latter would then quickly dilute with respect to the former, this hardly leaves us with any bouncing scenarios of relevance to our Universe. 
A caveat is that while a comoving observer sees diverging structures, there exists at least the mathematical possibility of some different gauge with less singular 
properties (for a case study of another type of first-order modified gravity, see \cite{Koivisto:2010jj}). 

\section{On generalisations to $f(T,B)$ and other models}
\label{generalisations}

Since the torsion tensor components from previous sections are independent of a particular model as long as the simplest FRW ansatz is a solution, one can easily generalise our discussion to any other teleparallel model at hand if equations of motion are known in a convenient form. Let us give some details to the latter.

The first step to make is to observe a very nice relation for a variation of the connection coefficient
\begin{equation}
\label{connvar}
\delta \Gamma^{\alpha}_{\mu\nu}=\bigtriangledown_{\mu}\left(e^{\alpha}_B \delta e^B_{\nu}\right)=\mathop{\bigtriangledown_{\mu}}\limits^{(0)}\left(e^{\alpha}_B \delta e^B_{\nu}\right)+K^{\alpha}_{\hphantom{\alpha}\mu\beta}e^{\beta}_B \delta e^{B}_{\nu}-K^{\beta}_{\hphantom{\beta}\mu\nu}e^{\alpha}_B \delta e^{B}_{\beta}
\end{equation}
which can be proven by direct computation of $\bigtriangledown_{\mu}\left(e^{\alpha}_B \delta e^B_{\nu}\right)$ and comparison with $\delta\left(e^{\alpha}_B \partial_{\mu} e^{B}_{\nu}\right)$. Moreover, as can be easily seen, it is also valid with arbitrary spin connection provided that the variation is performed over the tetrad with the spin connection kept fixed.

For the $f(T,B)$ models with $B=2\mathop{\bigtriangledown_{\mu}}\limits^{(0)}T^{\mu}$ the full procedure for deriving equations of motion is contained in our approach to $f(T)$. Indeed, the variation is given by
$$\delta S=\int\left(f\delta \|e\|-\|e\|f_T \delta\mathop{R}\limits^{(0)}+\|e\| (f_B - f_T)\delta B \right).$$
The variation of the Ricci scalar is well-known from GR, while for the $B$-term we have
$$\delta B=2T^{\mu}\partial_{\mu}\left(e^{\nu}_B\delta e^{B}_{\nu}\right)+2\mathop{\bigtriangledown_{\mu}}\limits^{(0)}\delta \left(g^{\mu\nu}T_{\nu}\right)$$
with the variation of the torsion vector derived from (\ref{connvar}) as
\begin{equation}
\label{vecvar}
\delta T_{\mu}=\partial_{\mu}\left(e^{\nu}_B\delta e^{B}_{\nu}\right)-\mathop{\bigtriangledown_{\alpha}}\limits^{(0)}\left(e^{\alpha}_B \delta e^B_{\mu}\right)-K^{\alpha}_{\hphantom{\alpha}\alpha\nu}e^{\nu}_B \delta e^{B}_{\mu}+K^{\nu}_{\hphantom{\nu}\alpha\mu}e^{\alpha}_B \delta e^{B}_{\nu}
\end{equation}
where $K^{\alpha}_{\hphantom{\alpha}\alpha\nu}=-T_{\nu}$.

After some simple algebra we get the equation of motion:
\begin{equation}
\label{TB}
f_{T}\mathop{G_{\mu\nu}}\limits^{(0)}+\frac12 \left(f-f_{T}{\mathbb T}-f_B B\right)g_{\mu\nu}+\left(g_{\mu\nu}\mathop{\square}\limits^{(0)}- \mathop{\bigtriangledown_{\mu}}\limits^{(0)}\mathop{\bigtriangledown_{\nu}}\limits^{(0)}\right)f_B+S_{\mu\nu\alpha}\partial^{\alpha}(f_T - f_B)=8\pi G\cdot\Theta_{\mu\nu}.
\end{equation}
We see that if $f_B\neq 0$ equations of motion are of higher order (4th in symmetric and 3rd order in the antisymmetric parts). If $f=f(T+B)$ it reduces to the case of $f(R)$. Moreover, the trace part of the symmetric equation gives a wave equation for $f_B$. If $f(T,B)=f_1(T)+f_2(B)$, one can go for the same reduction of order as in $f(R)$ gravity.

Note that we have written equation (\ref{TB}) in such a way that all these terms remain intact if one adds more arguments to the function $f.$ One option is to have $f(T,B,\varphi,X,Y)$ with $X\equiv\frac12 g^{\mu\nu}(\partial_{\mu}\varphi)(\partial_{\nu}\varphi)$ and $Y\equiv T^{\mu}\partial_{\mu}\varphi$. In the terms which are already present in the equation (\ref{TB}) one would only need to understand the derivatives correctly, e.g. $\partial_{\mu}f_T=f_{TT} \partial_{\mu}{\mathbb T}+f_{TB} \partial_{\mu}B+f_{T\varphi}\partial_{\mu}\varphi+f_{TX} \partial_{\mu}X+f_{TY} \partial_{\mu}Y$, and the new terms to be added are
$$+\frac12\left(\mathop{\bigtriangledown_{\mu}}\limits^{(0)}\left(f_Y \partial_{\nu}\varphi\right)-\mathop{\bigtriangledown_{\alpha}}\limits^{(0)}\left(f_Y \partial^{\alpha}\varphi\right)g_{\mu\nu}+f_Y\left(K_{\mu\nu\alpha}\partial^{\alpha}\varphi - T_{\mu}\partial_{\nu}\varphi\right)-f_X (\partial_{\mu}\varphi)(\partial_{\nu}\varphi)\right)$$
in the left hand side of equation (\ref{TB}). We see that $Y$ contributes to the antisymmetric part of equations.

New GR models \cite{Hayashi:1979qx} require slightly more work. However, in case of looking for deviations from GR, it would also be nice to parametrise coefficients in front of independent scalars \cite{Bahamonde:2017wwk} in terms of deviations from $\mathbb T$. Then it would amount to adding even more arguments to our analysis. Yet, one may also consider non-minimal couplings to matter \cite{DAgostino:2018ngy}. 

\subsection{An example: scalar-torsion gravity}
\label{scalar-torsion}

For simplicity, we will consider in more detail only the extension to models $f({\mathbb T},\varphi)$ which, with the addition of the kinetic term $\sim \omega(\varphi)X$, then cover the many ''scalar-torsion gravity'' models whose covariant formulation was recently discussed in \cite{Hohmann:2018rwf} (see there for many earlier references). The action (\ref{action}) is then generalised to 
\begin{equation}
\label{action2}
S=-\int d^4 x  \| e\|\cdot \left[ f\left({\mathbb T},\varphi \right) + 2\omega(\varphi)X \right]\,.
\end{equation}
One may expect differences in the scalar sector of cosmological perturbations. We parameterise again the tetrad perturbations in the Newtonian gauge as in Section
\ref{parameterisation} and in addition let the scalar field fluctuate, denoting the perturbation as $\delta\varphi$. 
The scalar field equation of motion,
\begin{equation}
\Box \varphi + \frac{1}{\omega}\left( \omega_{,\varphi}X - \frac{1}{2}f_{,\varphi}\right) = 0\,, \nonumber
\end{equation}
is linearised into
\begin{eqnarray}
\delta\varphi'' & + & \left(2H+\frac{\omega_{,\varphi}}{\omega}\varphi'\right)\delta\varphi' - \left[\bigtriangleup + \left(\frac{\omega_{,\varphi}}{\omega}\right)_{,\varphi}X 
-   \left(\frac{f_{,\varphi}}{2\omega}\right)_{,\varphi} \right]\delta\varphi \nonumber \\
& = & \varphi'\phi'+\left(2\varphi''+4H\varphi'+2\frac{\omega_{,\varphi}}{\omega}X\right)\phi + 3\varphi'\psi' + \frac{1}{2\omega}f_{T\varphi}\delta {\mathbb T}\,. \nonumber
\end{eqnarray}
Let us then look at the antisymmetric part of the field equations,
$S_{[\mu\nu]\alpha}\partial^\alpha f_T=0$. We obtain now
\begin{equation}
\left[ 12a^{-2}f_{TT}\left(H'-H^2\right)H + f_{T\varphi}{\varphi}'\right]\psi - 4a^{-2}f_{TT}H^2\left(\bigtriangleup\zeta+3H\phi+3{\psi}'\right) + f_{T\varphi}H\delta\varphi=0\,. \nonumber
\end{equation}
In the case that $f_{TT}=0$ but $f_{T\varphi} \neq 0$, this equation becomes simply $\delta\varphi = -(\varphi'/H)\psi$. Thus the fluctuation of the scalar field is directly
proportional to the spatial Bardeen potential. We will restrict to this case in the following, and thus consider $f({\mathbb T},\varphi) = f_T(\varphi){\mathbb T} + 2V(\varphi)$. Furthermore,
by a redefinition of the field, $\varphi \rightarrow f_T(\varphi)$, we can without any essential loss of generality\footnote{The price to pay for the simplifications is just that the field $\varphi$ and 
the function $\omega$ are now considered in somewhat unconventional dimensions, $[\varphi]=[1/\omega]=[1/8\pi G]$.}
 fix $f({\mathbb T},\varphi) = \varphi {\mathbb T} + 2V(\varphi)$.
With this simplification, the field equations are 
\begin{equation}
\varphi \mathop{G_{\mu\nu}}\limits^{(0)} + S_{\mu\nu}{}^{\alpha}\varphi_{,\alpha} = \omega\left( \varphi_{,\mu}\varphi_{,\nu}-X g_{\mu\nu} \right) - V g_{\mu\nu} + \Theta_{\mu\nu}\,. \nonumber
\end{equation}
The background equations can be read off immediately:
\begin{eqnarray}
3 H^2\varphi & = & -\omega X + a^2\left( V + \rho\right)\,, \nonumber \\  
\left(2H' + H^2\right)\varphi   + 2H\varphi' & = & \omega X  + a^2\left( V - p\right)\,, \nonumber \\
\omega\left( \varphi'' + 2H\varphi'\right) - \omega_{,\varphi} X & = &  - 6H^2 - a^2V_{,\varphi}\,, \nonumber
\end{eqnarray}
where $X= -{\varphi'}{}^2/2$.
Let us consider the scalar-torsion system sourced by a perfect fluid.
The four independent linearised equations can then be summarised as follows. \newline
The energy constraint:
\begin{equation} \label{efe1}
2\varphi\left[\bigtriangleup \psi -3H\left(\psi'+H\phi\right)\right] = \omega\varphi'\left(\delta\varphi' - \varphi'\phi\right) - \left(3H^2 + \omega_{,\varphi}X - a^2V_{,\varphi}\right)\delta\varphi + 
a^2\rho\delta\,.
\end{equation}
The velocity propagation: 
\begin{equation} \label{efe2}
\varphi\left( \psi' + H\phi\right) +\frac{1}{2} \varphi'\psi = \frac{1}{2}\left( H + \omega\varphi' \right) \delta\varphi + \frac{1}{2}a^2(\rho+p)\delta \mathfrak{u}\,.
\end{equation}
The pressure constraint:
\begin{eqnarray}
\varphi\left[\psi'' + H\left( 2\psi'+\phi'\right) + \left(2H' + H^2\right)\phi  +   \frac{1}{3}\bigtriangleup\left(\phi-\psi\right)\right]  +  \varphi'\left(\frac{1}{3}\bigtriangleup\zeta + \psi' + 2H\phi\right)
\nonumber \\   =  
\left( \frac{1}{2}\varphi' \omega +H\right)\delta\varphi' - \frac{1}{2}\left(\omega_{,\varphi}X+V_{,\varphi}\right)\delta\varphi + X\omega\phi + \frac{1}{2}a^2\delta p \,.
\end{eqnarray}
The shear propagation:
\begin{equation}
\phi-\psi = -\frac{\varphi'}{\varphi}\zeta\,.
\end{equation}
Despite the presence of the kinetic term for the scalar $\varphi$, it does not propagate a new degree of freedom. This is because the additional constraint we had obtained from the 
antisymmetric field equation, $\delta\varphi = -(\varphi'/H)\psi$. If we restrict to vacuum, $\rho=p=0$, it is easy to see that the combined two first equations (\ref{efe1},\ref{efe2})
entail that $\phi=\psi=0$. 

Another case of interest is the evolution of dust ($\rho \neq 0$, $p=0$) perturbations at the quasistatic limit ($k^2 \gg H^2$). At this limit, 
the algebra analogous to that in Section \ref{comoving} becomes very simple and quickly shows that, as expected, all the perturbations of the tetrad, as well as the perturbation of the scalar field,
can be directly related to the matter overdensity $\delta$. We obtain:
\begin{equation} 
k^2\phi = -\left(\frac{ H^2\varphi+\omega X}{H^2}\right)\frac{a^2\rho\delta}{2\varphi}\,, \quad
k^2\psi =  -\frac{a^2\rho\delta}{2\varphi}\,, \quad
k^2\zeta =  -\left(\frac{\omega \varphi'}{2H^2\varphi}\right)\frac{a^2\rho\delta}{2\varphi}\,, \quad
k^2\delta\varphi =  \left(\frac{\varphi'}{H}\right)\frac{a^2\rho\delta}{2\varphi}\,. \nonumber
\end{equation}
The two first equations can be regarded as relativistic generalisations of the Poisson equation.
Combining these equations, we obtain a closed form evolution equation for the matter density, 
\begin{equation} \label{delta2}
\delta'' + H\delta' = \frac{a^2\rho}{2\varphi}\left( 1 + \frac{\omega X}{\varphi H^2}\right) \delta\,.
\end{equation}
This generalises the result (\ref{delta}), at the small-scale limit, to more general scalar-torsion theories. The effective Newton's constant
depends now also on the kinetic term of the field. When we neglect it, $\omega=0$, we consistently recover the result of the $f(T)$ models,
since in the scalar-torsion description of the $f(T)$ models we have that $\varphi = f_T$.

\section{On generalisations to curved FRW and rotated tetrads}
\label{nonflat}

In this Section we consider the cosmological perturbations in non-flat FLRW background. Regarding cosmological observations, this task appears less
urgent since all the cosmological data is consistent with the assumption that our Universe has flat spatial sections, which is usually understood as
a consequence of inflation in the very early Universe. 

However, there is theoretical interest to investigate the propagating modes of the teleparallel theory in
more non-trivial backgrounds. As we have again confirmed above, in the conformally flat and isotropic, homogeneous background we do not excite the extra degree of freedom that is expected to nevertheless be present in the theory \cite{Ferraro:2018tpu}. 
Rather than proceeding to nonlinear order in perturbation theory, one might instead to instead perturb the background solutions in different background 
geometry\footnote{In this regard, the static, spherically symmetric backgrounds might be promising since there appears a discontinuity of the solutions at the limit $f_{TT}\rightarrow 0$ \cite{Ruggiero:2015oka}.} in the hope getting the extra degree of freedom caught red-handed. However, in the following analysis we encounter an obstacle, which reveals another curious theoretical feature of the modified teleparallel gravity models, and makes it particularly clear that there is an issue with the proper parallelisation that needs to be resolved regardless of the background. We will first recall the classification of tetrads into the ''good'' and the ''bad'' ones, and then point out that a finer characteristics is required at the perturbative level. 

Let us have a quick look at the models in the curved FRW background in the isotropic coordinates written as
\begin{equation} \label{isotropic} \nonumber
ds^2 = -a^2(\tau)\left(-d\tau^2 +  X_i (dx^i)^2\right)\,, \quad X_i \equiv 1 + \frac{K(x^i)^2}{1+K\delta_{jk}x^jx^k}\,,
\end{equation}
where $K$ is the curvature parameter. With the most obvious diagonal choice of the tetrads, we obtain the torsion scalar 
\begin{equation} \nonumber
{\mathbb T} = 6\frac{H^2}{a^2} - \frac{6K^4 x^2y^2z^2}{a^2\left(1-Kx^2-Ky^2-Kz^2\right)}\left(1-Kx^2-Ky^2\right)^{-1}\left(1-Ky^2-Kz^2\right)^{-1}\left(1-Kx^2-Kz^2\right)^{-1}\,.
\end{equation} 
Obviously, unless $K = 0$, this scalar is not isotropic and homogeneous, contradicting the symmetries of the background. A general ${\mathbb T}$-dependent action won't thus allow any
non-trivial solutions for this background. This means that the diagonal tetrads are no ''good'' \cite{Tamanini:2012hg} i.e. not compatible with the zero spin connection as explained in \cite{Golovnev:2017dox}.

Following the original work \cite{Ferraro:2011us}, we now switch to the hyperspherical coordinates. For definiteness we consider a closed universe background, and 
again will work with the quasi-Newtonian gauge perturbations,
\begin{equation}
ds^2 = a^2(\tau)\left[-(1+2\phi)d\tau^2 + K^{-1}(1-2\psi)\gamma_{ij}dy^i dy^j  \right]\,, \nonumber
\end{equation}
where the curvature parameter $K>0$ and the metric of the 3-sphere is  
\begin{equation}
\gamma_{ij}dy^idy^j=d\Psi^2 + \sin^2{\Psi}\left(d\Theta^2 +\sin^2\Theta d\Phi^2\right)\,. \nonumber
\end{equation}
The tetrad perturbations will involve also the gradient of the $\zeta$, which reads in the $y^i$-coordinates as
\begin{equation} \label{grad}
\zeta_{;i}d y^i = \frac{\partial\zeta}{\partial \Psi}d\Psi + \frac{1}{\sin\Psi}\frac{\partial\zeta}{\partial \Theta}d\Theta
+ \frac{1}{\sin\Psi\sin\Theta}\frac{\partial\zeta}{\partial \Phi}d\Phi\,, 
\end{equation}
and the Laplacian operator $\bigtriangleup$ in the hyperspherical basis is
\begin{equation}
\gamma^{ij}\zeta_{;i;j} = \zeta_{,\Psi\Psi} + 2\cot{\Psi}\zeta_{,\Psi} + \csc^2{\Psi}\left(\zeta_{,\Theta\Theta} + 
\cot{\Theta}\zeta_{,\Theta} + \csc^2{\Theta}\zeta_{,\Phi\Phi}\right)\,. \label{laplacian}
\end{equation}

An obvious choice for the tetrads is again a diagonal one, which is the most direct generalisation of our previous choices. 
We write thus\footnote{The are other possible assignments for the $\zeta$-part which would lead to the same metric, such as $e^0_i  =  a(\tau)\sqrt{\gamma^{ij}}\zeta_{;j}$, 
$e^a_0  =  a(\tau)\zeta_{;j}$. However, it is important to note that the $\zeta$-perturbation has inherited its derivative from the Lorentz basis, whereas
the $\beta$ (which our gauge sets to $\beta=\zeta$) is differentiated with respect to the coordinate indices. Other assignments than the above would lead to spurious coordinate-dependence in $\mathbb{T}$ and the field equations.} 
\begin{eqnarray}
e^\emptyset_0 & = & a(\tau)\cdot\left(1+\phi\right)\,, \nonumber \\
e^\emptyset_i & = & a(\tau)\cdot \zeta_{,i}\,,  \nonumber \\
e^a_0 & = & a(\tau)\cdot \delta^{ai}\zeta_{;i}\,,  \nonumber \\
e^a_j & = & a(\tau)\cdot \delta^{ai}\sqrt{K\gamma_{ij}}\left(1-\psi\right)\,.  \nonumber
\end{eqnarray}
However, again these diagonal tetrads turn out to be ''bad''. Therefore we should rotate them. The rotation we perform is specified by three Euler angles such that first we turn around the $\Psi$-axis by the angle
$-\arcsin({\cos{\Psi}})$, then around the $\Theta$-axis by the angle
$\pi/2-\Theta$, and finally around the $\Phi$-axis by the angle $\Phi$. 
The tetrad we then obtain seems like a ''good'' one (from the result (\ref{torsionscalar2}) we'll shortly arrive at) though a lot messier (we omit writing down the 16 nonzero components explicitly). At this point, it is more convenient to rescale the hyper-angular coordinates $y^i \rightarrow  \sqrt{K}y^i$, so that they have the conventional dimension of length. 

The components of the torsion become 
\begin{eqnarray}
T_{0ij}   & = & T_{0ij}^\zeta \,, \nonumber \\
T_{00j} & = & a^2 \left(\phi_{;j} - \zeta'_{,j} \right)\,, \nonumber \\
T_{ijk} - T_{ijk}^\zeta & = & a^2\left[2 \sqrt{K}\sin^2{\left( \frac{\Psi}{\sqrt{K}}\right)}\sin{\left(\frac{\Theta}{\sqrt{K}}\right)}\epsilon_{ijk}\left(1-2\psi\right)  + \gamma_{ij} \psi_{,k}-\gamma_{ik}\psi_{,j}\right]\,, \nonumber \\
T_{i0j} -T_{i0k}^\zeta & = & a^2\gamma_{ij}\left( H - 2H\psi-\psi^{\prime}\right)\,, \nonumber
\end{eqnarray}
where the contribution from the extra perturbation assumes now a more complicated form\footnote{For concreteness, when setting $K=1$, the components of $T_{i0k}^\zeta$ look like 
$T_{\Psi 0 \Psi}^\zeta  =  -a^2\zeta_{,\Psi\Psi}$, $T_{\Psi 0 \Theta}^\zeta  =  -a^2\left( \zeta_{,\Psi\Theta} +\cot{\Psi}\zeta_{,\Theta} - \csc{\Theta}\zeta_{,\Phi}\right)$, 
$T_{\Psi 0 \Phi}^\zeta  =  -a^2\left( \zeta_{,\Psi\Phi} - \sin{\Theta}\zeta_{,\Theta} - \cot{\Psi}\zeta_{,\Phi} \right)$, $T_{\Theta 0 \Psi}^\zeta  =  -a^2\left(\zeta_{,\Psi\Theta} - \csc{\Psi}\zeta_{,\Psi}  - \cot{\Theta}\zeta_{,\Phi}\right)$, etc. etc.} for other than the $T_{00j}$ components. 
However, the torsion scalar is neatly summed up into
\begin{equation} \label{torsionscalar2}
{\mathbb T}=\frac{6\left(H^2-K\right)}{a^2} -\frac{12H}{a^2}\left(H\phi+\psi^{\prime} + \frac{1}{3}\gamma^{ij}\zeta_{;i;j}\right) -\frac{12 K}{a^2}\psi\,. 
\end{equation}
Weird things happen in the antisymmetric sector. There are now two non-trivial equations. The three time-space components result in the same equation as previously, recall the solution (\ref{zeta}), taking into account the eigenvalues of the operator $\bigtriangleup=-k^2-3K$ in a closed universe. Explicitly, we get    
\begin{equation}
\label{zeta2}
\left(\frac{1}{3}k^2+K\right)\zeta=\psi^{\prime}-H\phi+\left(\frac{H^{\prime}}{H}+H\right)\psi\,. \nonumber
\end{equation}
The three antisymmetric spatial field equations now dictate that in addition we have to satisfy the constraint
\begin{equation}
\left(\frac{1}{3}k^2+H^2-H'\right)\zeta = \psi' + H\phi + \frac{K}{H}\psi \,. \nonumber
\end{equation}  
Obviously, the system is overconstrained. At the level of structure formation, the generalised models are not viable. 
Note that this conclusion remains in the limit $K\rightarrow 0$. It demonstrates that the rotated tetrads may remain ''good'' but become physically 
inequivalent. This of course reflects the fact that the $f(T)$ models break the rotational invariance (when the spin connection is neglected \cite{Golovnev:2017dox}).  

M. Hohmann has reported that differently boosted tetrads which produce the same spatially curved FRW metric can lead to already inequivalent background field equations in the $f(T)$ models \cite{hohmann}. 
In fact, it had been recently discovered, in the more general context of McVittie solutions, that in a Newman-Penrose basis there exist tetrads which accommodate the standard GR solutions for 
the flat FRW metric into the general $f(T)$ models \cite{Bejarano:2017akj} . One tool to study the inequivalence of solutions is given by the condition for the existence of remnant symmetries in a given background \cite{Ferraro:2014owa}, since obviously the transformations that change the physical predictions cannot be amongst the remnant symmetries. From the covariant perspective, those transformations will correspond to the degrees of freedom in the inertial connection \cite{Golovnev:2017dox}. 

The question arises whether the conclusion of the previous Sections that there are no extra degrees of freedom in the modified models, depends upon the choice of the tetrad we happened to make there. With tetrads in the Lorentz frame of Section \ref{parameterisation}, we obtained one, with above-chosen rotated frame we obtained two constraints. One may speculate on the existence of yet another frame where not one but two of the constraints would be eliminated, and the extra perturbation $\zeta$ would be unleashed to propagate\footnote{It remains to be clarified whether all of the field equations can always made be independent of the angular dependencies outside the perturbation variables i.e. statistically isotropic. This may require a perturbation-dependent rotation or boost, and perhaps be technically facilitated by considering the tangent space metric in an unconventional basis. We also leave for future investigations the possibility, which arises in the cases wherein the overall angular dependence does not cancel out at perturbative order, to interpret it as a (pseudo-)rotation of our observer frame with respect to the cosmic microwave background that could explain the anomalies that exist in the data according to e.g. \cite{Schwarz:2015cma}.}. In any case, it is clear that 1) ''good'' choices of tetrads corresponding to a given metric can be physically inequivalent and 2) the ''goodness'' at the background level can be lost by small perturbations. To recapitulate, 1) was already dramatically demonstrated in the results of \cite{Bejarano:2017akj}, and the problem we raise here due to 2) is the ambiguity of the propagating degrees of freedom
in modified teleparallel gravity models.

Let us mention that in the theory of Coincident GR \cite{BeltranJimenez:2017tkd} and its symmetric teleparallel modifications, such ambiguities may not exist, since this framework presents a physical rationale to determine the canonical frame (and the canonical coordinate system) \cite{Koivisto:2018aip,BeltranJimenez:2017tkd}. On the other hand, these ambiguities only demonstrate that already classically, the richness of solutions of a general tetrad theory cannot be captured within a solely metric description. From a different point of view, the ''unambiguous'' case of $f \sim \mathbb{T}$ 
has been criticised for the very opposite reason, namely that there is no way to distinguish between the infinite number of different Lorentz frames \cite{Garecki:2010jj}.

\section{Conclusions}
\label{conclusions}

We analysed cosmological perturbations in teleparallel $f(T,\varphi)$ models of gravity. Taking carefully into account the 16 independent perturbative components in a generic tetrad, classified into scalar, pseudoscalar, vector, pseudovector and tensor perturbations, we confirmed the previous conclusion of the absence of extra degrees of freedom in the flat FRW background in $f(T)$ models. This conclusion was generalised to the scalar-torsion models, which may appear surprising due to the explicit kinetic term that is added into the action (\ref{action2}). 

Other new results in this paper include the exact Poisson equation (\ref{poisson}) and the exact evolution equation for dust perturbations (\ref{delta}), where the latter is a special case of the second order differential equation (\ref{bardeen}) that makes no assumptions about the cosmological sources. The quasistatic equation governing the evolution of the matter spectrum was given also for the scalar-torsion theories as (\ref{delta}), which now allows to easily include the structure formation constraints when confronting these models with the available (e.g. SDSS \cite{Tegmark:2003uf}) and forthcoming (e.g. Euclid \cite{Amendola:2016saw}, SKA \cite{Carilli:2004nx}) cosmological precision data.

We also checked the behaviour of the linear perturbations at the critical turnover points in bouncing and recollapsing cosmologies. We deduced from (\ref{bounce}) that homogeneous
perturbations are divergent unless the action is contrived such that $f_{TT}(0)=0$, which rules out realistic bounces. 

The scalar-torsion case scrutinised in Section \ref{scalar-torsion} is a simple example amongst the various more general teleparallel 
modified gravity actions that have been proposed in the literature during the past few years. In Section \ref{generalisations} we sketched the general recipe to obtain the linearised cosmological equations for  various classes of models. We believe this will facilitate the more extensive analysis that is necessary to carry out for each of the models to sort out the viable ones according to their degrees of freedom and to understand their implications to cosmology beyond the time-dependence of the scale factor.

Finally, we pointed out a new feature of the perturbation system, which occurred with tetrads that were rotated to be compatible with spatial curvature. There was a discrepancy in the predictions of the modified models at the limit $K \rightarrow 0$. The conclusions can crucially depend upon the tetrad that is chosen to represent a given metric, spatially curved or otherwise. That is intriguing and calls for further studies.

\paragraph{Acknowlegdments.} The authors are grateful to the organisers of the TeleGrav2018 workshop in Tartu, and AG acknowledges Saint Petersburg State University travel grant 27801255 which made his participation possible. We thank Yi-Fu Cai, Stefano Camera and especially the authors of \cite{Bejarano:2017akj} for useful comments on the manuscript. 

\bibliography{telepert}

\begin{thebibliography}{10}

\bibitem{Aldrovandi:2013wha}
R.~Aldrovandi and J.~G. Pereira, {\em {Teleparallel Gravity}}, vol.~173.
\newblock Dordrecht: Springer, 2013.

\bibitem{Goenner:2004se}
H.~F.~M. Goenner, ``{On the history of unified field theories},'' {\em Living
  Rev. Rel.}, vol.~7, p.~2, 2004.

\bibitem{MOLLER1961118}
C.~M{\o}ller, ``Further remarks on the localization of the energy in the
  general theory of relativity,'' {\em Annals of Physics}, vol.~12, no.~1,
  pp.~118 -- 133, 1961.

\bibitem{Zlosnik:2018qvg}
T.~Złośnik, F.~Urban, L.~Marzola, and T.~Koivisto, ``{Spacetime and dark
  matter from spontaneous breaking of Lorentz symmetry},'' 2018.

\bibitem{moller1978crisis}
C.~M{\o}ller, {\em On the Crisis in the Theory of Gravitation and a Possible
  Solution}.
\newblock Kongelige Danske Videnskabernes Selskab: Matematisk-fysiske
  meddelelser, Munksgaard, 1978.

\bibitem{Koivisto:2018aip}
T.~Koivisto, ``{On an integrable geometrical foundation of gravity},'' {\em
  International Journal of Geometric Methods in Modern Physics}, vol.~15,
  p.~1840006, 2018.

\bibitem{Cai:2015emx}
Y.-F. Cai, S.~Capozziello, M.~De~Laurentis, and E.~N. Saridakis, ``{f(T)
  teleparallel gravity and cosmology},'' {\em Rept. Prog. Phys.}, vol.~79,
  no.~10, p.~106901, 2016.

\bibitem{Ferraro:2006jd}
R.~Ferraro and F.~Fiorini, ``{Modified teleparallel gravity: Inflation without
  inflaton},'' {\em Phys. Rev.}, vol.~D75, p.~084031, 2007.

\bibitem{Ferraro:2008ey}
R.~Ferraro and F.~Fiorini, ``{On Born-Infeld Gravity in Weitzenbock
  spacetime},'' {\em Phys. Rev.}, vol.~D78, p.~124019, 2008.

\bibitem{Bengochea:2008gz}
G.~R. Bengochea and R.~Ferraro, ``{Dark torsion as the cosmic speed-up},'' {\em
  Phys. Rev.}, vol.~D79, p.~124019, 2009.

\bibitem{Linder:2010py}
E.~V. Linder, ``{Einstein's Other Gravity and the Acceleration of the
  Universe},'' {\em Phys. Rev.}, vol.~D81, p.~127301, 2010.
\newblock [Erratum: Phys. Rev.D82,109902(2010)].

\bibitem{Geng:2011aj}
C.-Q. Geng, C.-C. Lee, E.~N. Saridakis, and Y.-P. Wu, ``{“Teleparallel”
  dark energy},'' {\em Phys. Lett.}, vol.~B704, pp.~384--387, 2011.

\bibitem{Jarv:2015odu}
L.~Jarv and A.~Toporensky, ``{General relativity as an attractor for
  scalar-torsion cosmology},'' {\em Phys. Rev.}, vol.~D93, no.~2, p.~024051,
  2016.

\bibitem{Hohmann:2018rwf}
M.~Hohmann, L.~Järv, and U.~Ualikhanova, ``{Covariant formulation of
  scalar-torsion gravity},'' {\em Phys. Rev.}, vol.~D97, no.~10, p.~104011,
  2018.

\bibitem{Kofinas:2014owa}
G.~Kofinas and E.~N. Saridakis, ``{Teleparallel equivalent of Gauss-Bonnet
  gravity and its modifications},'' {\em Phys. Rev.}, vol.~D90, p.~084044,
  2014.

\bibitem{Bahamonde:2015zma}
S.~Bahamonde, C.~G. Böhmer, and M.~Wright, ``{Modified teleparallel theories
  of gravity},'' {\em Phys. Rev.}, vol.~D92, no.~10, p.~104042, 2015.

\bibitem{Bahamonde:2017wwk}
S.~Bahamonde, C.~G. Böhmer, and M.~Krššák, ``{New classes of modified
  teleparallel gravity models},'' {\em Phys. Lett.}, vol.~B775, pp.~37--43,
  2017.

\bibitem{Cardone:2012xq}
V.~F. Cardone, N.~Radicella, and S.~Camera, ``{Accelerating f(T) gravity models
  constrained by recent cosmological data},'' {\em Phys. Rev.}, vol.~D85,
  p.~124007, 2012.

\bibitem{Krssak:2015oua}
M.~Krššák and E.~N. Saridakis, ``{The covariant formulation of f(T)
  gravity},'' {\em Class. Quant. Grav.}, vol.~33, no.~11, p.~115009, 2016.

\bibitem{Golovnev:2017dox}
A.~Golovnev, T.~Koivisto, and M.~Sandstad, ``{On the covariance of teleparallel
  gravity theories},'' {\em Class. Quant. Grav.}, vol.~34, no.~14, p.~145013,
  2017.

\bibitem{Li:2011wu}
B.~Li, T.~P. Sotiriou, and J.~D. Barrow, ``{Large-scale Structure in f(T)
  Gravity},'' {\em Phys. Rev.}, vol.~D83, p.~104017, 2011.

\bibitem{Nunes:2018xbm}
R.~C. Nunes, ``{Structure formation in $f(T)$ gravity and a solution for $H_0$
  tension},'' {\em JCAP}, vol.~1805, no.~05, p.~052, 2018.

\bibitem{Izumi:2012qj}
K.~Izumi and Y.~C. Ong, ``{Cosmological Perturbation in f(T) Gravity
  Revisited},'' {\em JCAP}, vol.~1306, p.~029, 2013.

\bibitem{Li:2018ixg}
C.~Li, Y.~Cai, Y.-F. Cai, and E.~N. Saridakis, ``{The effective field theory
  approach of teleparallel gravity, $f(T)$ gravity and beyond},'' 2018.

\bibitem{BeltranJimenez:2017tkd}
J.~Beltran~Jimenez, L.~Heisenberg, and T.~Koivisto, ``{Coincident General
  Relativity},'' 2017.

\bibitem{Hayashi:1979qx}
K.~Hayashi and T.~Shirafuji, ``{New General Relativity},'' {\em Phys. Rev.},
  vol.~D19, pp.~3524--3553, 1979.
\newblock [,409(1979)].

\bibitem{So:2006pm}
L.~L. So and J.~M. Nester, ``{On source coupling and the teleparallel
  equivalent to GR},'' in {\em {On recent developments in theoretical and
  experimental general relativity, gravitation, and relativistic field
  theories. Proceedings, 10th Marcel Grossmann Meeting, MG10, Rio de Janeiro,
  Brazil, July 20-26, 2003. Pt. A-C}}, 2006.

\bibitem{Tamanini:2012hg}
N.~Tamanini and C.~G. Boehmer, ``{Good and bad tetrads in f(T) gravity},'' {\em
  Phys. Rev.}, vol.~D86, p.~044009, 2012.

\bibitem{Hohmann:2017jao}
M.~Hohmann, L.~Jarv, and U.~Ualikhanova, ``{Dynamical systems approach and
  generic properties of $f(T)$ cosmology},'' {\em Phys. Rev.}, vol.~D96, no.~4,
  p.~043508, 2017.

\bibitem{Cai:2018rzd}
Y.-F. Cai, C.~Li, E.~N. Saridakis, and L.~Xue, ``{$f(T)$ gravity after GW170817
  and GRB170817A},'' {\em Phys. Rev.}, vol.~D97, no.~10, p.~103513, 2018.

\bibitem{Carilli:2004nx}
C.~L. Carilli and S.~Rawlings, ``{Science with the Square Kilometer Array:
  Motivation, key science projects, standards and assumptions},'' {\em New
  Astron. Rev.}, vol.~48, p.~979, 2004.

\bibitem{Camera:2013bwa}
S.~Camera, V.~F. Cardone, and N.~Radicella, ``{Detectability of Torsion Gravity
  via Galaxy Clustering and Cosmic Shear Measurements},'' {\em Phys. Rev.},
  vol.~D89, p.~083520, 2014.

\bibitem{Amendola:2016saw}
L.~Amendola {\em et~al.}, ``{Cosmology and fundamental physics with the Euclid
  satellite},'' {\em Living Rev. Rel.}, vol.~21, no.~1, p.~2, 2018.

\bibitem{Wu:2012hs}
Y.-P. Wu and C.-Q. Geng, ``{Matter Density Perturbations in Modified
  Teleparallel Theories},'' {\em JHEP}, vol.~11, p.~142, 2012.

\bibitem{delaCruz-Dombriz:2018nvt}
{\'A}.~de~la Cruz-Dombriz, G.~Farrugia, J.~L. Said, and
  D.~S{\'a}ez-Chillón~G{\'o}mez, ``{Cosmological bouncing solutions in
  extended teleparallel gravity theories},'' {\em Phys. Rev.}, vol.~D97,
  no.~10, p.~104040, 2018.

\bibitem{Battefeld:2014uga}
D.~Battefeld and P.~Peter, ``{A Critical Review of Classical Bouncing
  Cosmologies},'' {\em Phys. Rept.}, vol.~571, pp.~1--66, 2015.

\bibitem{Koivisto:2010jj}
T.~S. Koivisto, ``{Bouncing Palatini cosmologies and their perturbations},''
  {\em Phys. Rev.}, vol.~D82, p.~044022, 2010.

\bibitem{DAgostino:2018ngy}
R.~D'Agostino and O.~Luongo, ``{Growth of matter perturbations in non-minimal
  teleparallel dark energy},'' 2018.

\bibitem{Ferraro:2018tpu}
R.~Ferraro and M.~J. Guzmán, ``{Hamiltonian formalism for f(T) gravity},''
  {\em Phys. Rev.}, vol.~D97, no.~10, p.~104028, 2018.

\bibitem{Ruggiero:2015oka}
M.~L. Ruggiero and N.~Radicella, ``{Weak-Field Spherically Symmetric Solutions
  in $f(T)$ gravity},'' {\em Phys. Rev.}, vol.~D91, p.~104014, 2015.

\bibitem{Ferraro:2011us}
R.~Ferraro and F.~Fiorini, ``{Non trivial frames for f(T) theories of gravity
  and beyond},'' {\em Phys. Lett.}, vol.~B702, pp.~75--80, 2011.

\bibitem{hohmann}
M.~Hohmann and {\it et al}, ``{“Cosmological” tetrads and spin connections
  in teleparallel gravity},'' 2018.

\bibitem{Bejarano:2017akj}
C.~Bejarano, R.~Ferraro, and M.~J. Guzmán, ``{McVittie solution in f(T)
  gravity},'' {\em Eur. Phys. J.}, vol.~C77, no.~12, p.~825, 2017.

\bibitem{Ferraro:2014owa}
R.~Ferraro and F.~Fiorini, ``{Remnant group of local Lorentz transformations in
  $\mathcal{f}(T)$ theories},'' {\em Phys. Rev.}, vol.~D91, no.~6, p.~064019,
  2015.

\bibitem{Schwarz:2015cma}
D.~J. Schwarz, C.~J. Copi, D.~Huterer, and G.~D. Starkman, ``{CMB Anomalies
  after Planck},'' {\em Class. Quant. Grav.}, vol.~33, no.~18, p.~184001, 2016.

\bibitem{Garecki:2010jj}
J.~Garecki, ``{Teleparallel equivalent of general relativity: A Critical
  review},'' in {\em {Hypercomplex Seminar 2010: (Hyper)Complex and
  Randers-Ingarden Structures in Mathematics and Physics Bedlewo, Poland, July
  17-24, 2010}}, 2010.

\bibitem{Tegmark:2003uf}
M.~Tegmark {\em et~al.}, ``{The 3-D power spectrum of galaxies from the
  SDSS},'' {\em Astrophys. J.}, vol.~606, pp.~702--740, 2004.

\end{thebibliography}
\bibliographystyle{ieeetr}

\end{document}